# Upper critical field, superconducting energy gaps, and Seebeck coefficient in $La_{0.8}Th_{0.2}OFeAs$

J. Prakash[a], S. J. Singh[b], S. Patnaik[b]* and A. K. Ganguli[a]*

[a]Department of Chemistry, Indian Institute of Technology, New Delhi 110016 India

[b]School of Physical Sciences, Jawaharlal Nehru University, New Delhi 110067 India

# Abstract

We report the synthesis and characterization of a new electron-doped La-oxypnictide superconductor by partial substitution of lanthanum by thorium. The superconducting transition temperature at about 30.3 K was observed in $La_{0.8}Th_{0.2}OFeAs$ which is the highest in La-based oxypnictide superconductors synthesized at ambient pressure. We find that the decrease in lattice parameters with Th doping in LaOFeAs is more drastic as compared to that obtained by high pressure (6 GPa) synthesis of oxygen deficient LaOFeAs. The resistivity and magnetic susceptibility measurements yield an upper critical field $H_{c2}$ (0) of 47 T. Partial substitution of Th in place of La induces electron doping as evidenced by negative Seebeck coefficient. The temperature dependent magnetic penetration depth data provides strong evidence for a nodeless low energy gap of 1.4 meV.

# Introduction:

The recent discovery of superconductivity in layered rare-earth transition metal (Ln) oxypnictides of the type LnOFeAs, has given rise to immense interest in the field of high temperature superconducting materials. Kamihara et al. first showed that the iron based compound LaOFeAs exhibits superconductivity with transition temperature 26 K by substitution of oxygen by fluorine partially [1]. Detailed band structure calculations in this novel class of superconductors indicate that the ground state of these materials corresponds essentially to a low carrier density metal with disconnected Fermi surfaces and in close proximity to an itinerant magnetic state, particularly a spin density wave (SDW) correlation at around 150 K. It is proposed that superconductivity is induced by the suppression of the SDW state through electron doping [2-5]. This has lead to $T_c$ being increased up to 55 K with replacement of La by smaller rare-earths Ce, Sm, Nd, Pr and Gd [6-10]. These are the first non copper material that exhibit superconductivity above 50 K.

The oxypnictides LnOFeAs crystallize in the layered tetragonal ZrCuSiAs structure (space group: *P4/nmm*) with alternating layers of La-O and Fe-As. The ionic La-O layer acts as a charge reservoir and Fe-As layer acts as charge carrier [11]. We have earlier reported the potassium fluoride doped La-based oxypnictide with nominal composition $La_{0.8}K_{0.2}O_{0.8}F_{0.2}FeAs$ which show superconducting transition around 26.5 K [12]. The above superconductors are mainly induced by electron-doping by substitution of $F^-$ ion in place of $O^{-2}$. There have also been reports [13] of Co-doping at Fe sites to induce superconductivity. Recently Wang et al. have established another route for electron doping, e.g., substitution of $Th^{+4}$ in place at the Ln (rare earth) site in LnOFeAs

(Ln = Gd, Tb). This has lead to superconductivity at 56, 50 and 38 K in (Gd/Th)OFeAs, (Tb/Th)OFeAs and (Nd/Th)OFeAs respectively [14-16]. In this paper, we have investigated the effect of thorium doping at the lanthanum site in (LaOFeAs) on its electrical and magnetic properties. The spin density wave state is completely suppressed upon Th doping and superconducting transition is observed at 30.3 ($\pm$0.05) K in La$_{0.8}$Th$_{0.2}$OFeAs which is the highest T$_c$ reported in La based oxypnictides synthesized at ambient temperature. We also report our investigation on La and O deficient compounds of the type La$_{1-\delta}$OFeAs and LaO$_{1-\delta}$FeAs ($\delta \leq 0.2$). The thermodynamic parameters such as the upper critical field and coherence length have been estimated. The temperature dependence of rf penetration depth exhibits strong evidence in support of an exponential isotropic pairing mechanism.

## Experiment:

Polycrystalline samples with nominal compositions of La$_{1-x}$Th$_x$OFeAs(x= 0.1 and 0.2), La$_{1-\delta}$OFeAs ($\delta$=0.1 and $\delta$=0.2) and LaO$_{1-\delta}$FeAs ($\delta$=0.1 and $\delta$=0.2) were synthesized by conventional solid state reaction method [12, 17] using high purity La, La$_2$O$_3$, ThO$_2$ and FeAs. FeAs was obtained by heating Fe and As chips in evacuated sealed silica tubes at 800 °C for 24 h. The reactants were taken in stoichiometric ratio and sealed in evacuated silica (10$^{-4}$ torr) tubes and heated at 950 °C for 24 hours. The resulting powder was ground and compacted into disks. The disks were wrapped in Ta foil and sealed in evacuated silica ampoules. The compacted disks were sintered at 1180 °C for 48 hours and then cooled to room temperature. All the materials were handled in a nitrogen filled glove box. The samples were characterized by powder X-ray diffraction (XRD) using Cu-$K\alpha$ radiation. The lattice parameters were obtained from a least square fit to the observed $d$ values.

The resistive and magnetic measurements were carried out using a cryogenic 8 T cryogen-free magnet in conjunction with a variable temperature insert (VTI). The electrical properties were studied by standard four probe technique. Contacts were made using 44 gauge copper wires with air drying conducting silver paste. For magnetic measurements, the magnetic field (0-5 T) was applied perpendicular to the probe current direction and the data were recorded during the warming cycle with heating rate of 1 K/min. The inductive part (real part) of the magnetic susceptibility was measured via a tunnel diode based rf (2.3MHz) penetration depth technique [18] attached to this cryogen-free magnet system. A change in magnetic state of the sample results in a change in the inductance of the coil and is reflected as a shift in the oscillator frequency which is measured by an Agilent 53131A counter. Energy dispersive analysis by X- rays (EDX) was carried out on sintered pellets of the compounds on a Zeiss electron microscope in conjunction with a BRUKER EDX system. The thermo electric power (TEP) data were obtained in the bridge geometry across a 2 mm by 3 mm platelet.

**Results and discussion:**

Figure 1(i) shows the powder x-ray diffraction patterns of the sample with nominal compositions $La_{1-x}Th_xOFeAs$ (x= 0.1 and 0.2), $La_{1-\delta}OFeAs$ ($\delta$=0.1 and $\delta$=0.2) and $LaO_{1-\delta}FeAs$ ($\delta$=0.1 and $\delta$=0.2). The majority of the observed reflections for all the above mentioned compounds could be indexed satisfactorily on the basis of tetragonal ZrCuSiAs-type structure with a secondary minor phase of $ThO_2$ in case of Th doped samples. The refined lattice parameters for the $La_{1-\delta}OFeAs$ ( **a** = 4.0279(8)Å, **c** = 8.731(4) Å for $\delta$=0.1 and **a** = 4.0265(9)Å, **c**= 8.728(3) Å for $\delta$=0.2) and $LaO_{1-\delta}FeAs$(**a**= 4.0286(9) Å, **c**= 8.734(5) Å for $\delta$=0.1 and **a**= 4.0243(8) Å, **c**= 8.722(3) Å for $\delta$=0.2) were found to be smaller than the parent LaOFeAs [19] (a= 4.038 Å, c= 8.753 Å) which is expected since the lanthanum and oxygen deficiency leads to decrease in the lattice parameters.

The refined lattice parameters (**a** = 4.022(1) Å and **c** = 8.729(4) Å for x=0.1 and **a** = 4.018(1) Å and **c** = 8.713(6) Å for x=0.2) for Th doped LaOFeAs were smaller than those reported for pure LaOFeAs which is due to the incorporation of the smaller thorium at lanthanum sites. A minor amount of $ThO_2$ can be seen in the powder XRD pattern (figure 1(i)). It should be noted that the earlier reports of Th doped GdOFeAs, TbOFeAs and NdOFeAs superconductors also report the presence of $ThO_2$ as a secondary phase. It is to be noted that the **a** lattice parameter in $La_{0.8}Th_{0.2}OFeAs$ is lower than that reported for the oxygen deficient $LaO_{0.85}FeAs$ (**a**= 4.022(2) Å and **c**= 8.707(1) Å) superconductor ($T_c$=31.2 K) obtained under high pressure [20] which results in increase of chemical pressure on the Fe-As plane. The study of Ren et al. indicated that non-fluorine oxypnictide superconductors could be obtained by using high pressure(6GPa) synthesis and further increase in chemical pressure by substituting smaller ions in the Ln-O layers ( in turn affecting the Fe-As layers) leads to enhancement of $T_c$ [20]. In figure 1(ii), we show the energy dispersive X-ray (EDX) spectrum of $La_{0.8}Th_{0.2}FeAsO$ which confirms the presence of La, Th, Fe, As and O in the desired atomic ratio.

The zero field resistivity as a function of temperature is shown in figure 2 as measured by standard four probe method. Figure 2 (a) shows the temperature dependence of resistivity for the lanthanum and oxygen deficient LaOFeAs compounds along with the resistivity of $La_{0.9}Th_{0.1}OFeAs$. All the samples exhibit a sudden decrease of resistivity at ~ 150 K. The resistivity continues to decrease below 150 K and shows a minimum at ~ 70 K. However for $La_{0.9}Th_{0.1}OFeAs$ the reistivity becomes almost independent of temperature (below ~ 50 K). Ren et al [20] have earlier reported the synthesis of non-superconducting $LaO_{0.85}FeAs$ at ambient pressure which is similar to the behavior of the oxygen deficient oxypnictides $LaO_{1-\delta}FeAs$ ($\delta=0.1$ and $\delta=0.2$) prepared by us. These results confirm that the oxygen and lanthanum deficiency does not induce superconductivity in these La-based oxypnictides prepared at ambient pressure. Figure 2 (b) shows the zero resistivity measurement with respect to temperature for $La_{0.8}Th_{0.2}OFeAs$. Inset of this figure shows the variation of resistivity up to room temperature. The resistivity decreases monotonously with decreasing temperature and a rapid drop was observed at 30.3 K showing onset of superconductivity. The offset ($T_c(0)$) was found to be 26.7 K. The criteria used for determination of transition temperature has been reported earlier [17] and schematically shown in figure 2 (b). To confirm the presence of diamagnetic behavior, we measured the inductive part of susceptibility up to 35 K on the same sample. The inset of figure 3 shows the temperature dependence of the magnetic susceptibility. This attests the appearance of superconductivity in this polycrystalline sample. The broadening of the magnetic transition indicates a certain degree of inhomogeneity in the polycrystalline sample that also explains the observation of slightly lower magnetization onset $T_c$ (by ~ 1.8 K) as compared to the resistivity $T_c$. The inhomogeneity (primarily due to grain boundaries) of the sample is also clear from the value of the residual resistivity value (RRR = $\rho_{300} / \rho_{31}$) of 2.76. Nevertheless, we emphasize that the onset of superconductivity in $La_{0.8}Th_{0.2}OFeAs$ represents the highest transition temperature in La-based oxypnictides synthesized at ambient pressure.

To get more insight into the pairing mechanism in this multiband superconductor, we plot the shift in rf penetration depth frequency as a function of temperature (figure 3). This dependence is directly related to the anisotropy of the superconducting energy gaps. It is to be pointed out that the technique of rf penetration depth works better for sintered polycrystalline samples as compared to point contact or tunneling techniques [21], as the length scale probed here are two orders of magnitude higher for high $\kappa$ materials such as the oxypnictides. Further, complications arising out of phonon and magnetic impurities that dominate specific heat and thermal conductivity data are absent here [22].

The BCS expression for temperature dependence of penetration depth for an isotropic s-wave pairing is given by,

$$\Delta\lambda(T) = \lambda(0) \times \sqrt{\frac{\pi\Delta(0)}{2T}} \exp[-\Delta(0)/T] \qquad (1)$$

where, $\Delta\lambda(T)$ is difference between penetration depth at temperature T and at lowest measurement temperature of 1.8 K. $\lambda(0)$ and $\Delta(0)$ are the zero temperature values of penetration depth and energy gap respectively. In the tunnel diode oscillator technique, the change in penetration depth is related to measured shift in frequency, $\Delta\lambda = -G\ \Delta f$, where G is a geometrical constant that is calibrated to be ~ 10 for our set up. Several experiments have been reported to determine the superconducting gap in this new class of superconducting materials. Hunte et al. [23] have showed that $LaO_{0.9}F_{0.1}FeAs$ exhibits signs of two gap behavior similar to that in $MgB_2$. Lie Shan et al. [21] have reported the result of point contact spectroscopy (PCAS) which is best described for p-wave or d-wave gap with maximum gap of 3.9±0.7 meV. Mu Gang et al. [24] reported specific heat measurement for $LaO_{0.9}F_{0.1-\delta}FeAs$ which showed the maximum gap 3.4±0.5 meV based on d-wave mode. In figure 3 we show our data in the range $0<T/T_c<0.3$. We fitted the data for an isotropic s –wave gap and a d-wave power law ($T^2$) dependence. The solid line in figure 3 shows a fitting curve for an isotropic single gap model that evidently yields better fitting ($R^2 = 0.99$) to the experimental data according to equation (1) with the gap value of $\Delta_0/k_B = 16.6$ ( gap ~ 1.4 meV). We note that while fitting the data to Eq. 1 will yield the true estimate of the low lying gap, more information on the possibility of multiple gaps and anisotropy will need analysis of the superfluid density ( $\lambda^2(0)/\lambda^2(T)$). Since the sample has a transition temperature at 30.3 K, we calculate $\Delta_0/k_BT_c = 0.54$ and note that this value is smaller than the weak-coupling s-wave BCS value of 1.76. This is suggestive of a significant gap anisotropy or multiple gaps as in $MgB_2$ [25] ($\Delta_0/k_BT_c = 0.76$). The multi-gap behavior in La(O/F)FeAs has also been substantiated from the very high upper critical field indicated in high field transport measurements [23]. A universal two gap feature for La-based oxypnictide is also indicated from nuclear quadrupole measurements on $LaFeAsO_{0.92}F_{0.08}$ as reported by Kawasaki et al. [26]. In the inset we show the temperature variation of rf penetration depth for the entire superconducting temperature range.

To obtain the upper critical field ($H_{c2}$) we have studied the temperature dependence of the resistivity under different magnetic fields (figure 4). With increasing field, the transition temperature ($T_c$) shifts to lower values and the transition width gradually becomes broader, similar to the high $T_c$ cuprate superconductors [27]. This suggests the strong anisotropy of the critical field and possible propensity to thermal activated vortex dynamics [28]. The width of in- field superconducting transition indicates a broad region of flux-flow resistivity. Using the criteria of 90 % and 10 % of normal state resistivity ($\rho_n$), we obtained the upper critical field $H_{c2}$ and the irreversibility field $H^*(T)$ (field corresponding to suppression of bulk critical current density). The H-T phase diagram for sample ($La_{0.8}Th_{0.2}OFeAs$) is shown in inset of figure 4. We note that these are approximate estimations. Further, by using the Werthamer-Helfand-Hohenberg (WHH) formula $H_{c2}(0) = -0.693\, T_c\, (dH_{c2}/dT)_{T=T_c}$ [29], the zero field upper critical field $H_{c2}(0)$ can be calculated. A slope of -2.23 for $dH_{c2}/dT$ was estimated from the H–T phase diagram. Using $T_c$ = 30.3 K, we obtain $H_{c2}$ = 47 T. From the value of $H_{c2}(0)$, we can calculate the mean field Ginzberg-Landau coherence length $\xi = (\Phi_0 / 2\pi H_{c2}(0))^{1/2}$. Using $\Phi_0 = 2.07 \times 10^{-7}$ G cm$^2$ and the $H_{c2}(0)$ value, we obtain a coherence length ~ 26 Å. This value is higher than that reported for La(O/F)FeAs [12].

Since bulk superconductivity was realized by thorium doping in place of La, it is expected to be an electron doped superconductor. However, in a multiband scenario, both electron and hole pockets are possible. In order to confirm the dominant conduction mechanism, we have carried out studies on the temperature dependence of thermoelectric power (S) (figure 5). The thermoelectric power drops to zero sharply below the superconducting transition temperature. Also the negative value of the thermoelectric power indicates that its major carriers are electron like (n type), similar to that in La(O/F)FeAs superconductor [17] and is in contrast to the hole like carriers observed in (K/Sr)Fe$_2$As$_2$ [30]. At room temperature, the value of the thermopower is – 16 µV/K and shows a minimum at 90 K (S = - 73.1 µV/K). It is very interesting to note that the profile of S vs. T curve is similar to that of the low charge density metals like underdoped high T$_c$ cuprates except for the negative sign. We also note that the minima of the negative thermopower shifts to lower temperature with Th doping as compared to F doped LaOFeAs [17]. Further, the inset of figure 5 shows zero field resistivity above T$_c$ that exhibits a quadratic temperature dependence $\rho = \rho_0 + BT^2$ in the temperature range of T between 31 and 150 K. Here $\rho_0$ is the residual resistivity determined from the impurity scattering process. The T$^2$ dependence of $\rho$ below 150 K is reminiscent of the dominant electron-electron scattering process. From the experimental data, the values of $\rho_0$ = 1.72 mΩ-cm and B = 7.49 × 10$^{-5}$ mΩ-cm K$^{-2}$ were estimated. We note that the value of the slope B extracted here is larger than that reported value for La(O/F)FeAs [17] and both $\rho$(T) and S(T) suggest a stronger dominance of electron correlation effects in Th-doped superconductor.

In summary, we have synthesized three series of oxypnictides compounds with nominal compositions of La$_{1-x}$Th$_x$OFeAs(x= 0.1 and 0.2), La$_{1-\delta}$OFeAs($\delta$=0.1 and $\delta$=0.2) and LaO$_{1-\delta}$FeAs($\delta$=0.1 and $\delta$=0.2) by high temperature sealed tube method. Lanthanum and oxygen deficient compounds were found to be semimetallic like the parent LaOFeAs oxypnictides. A new superconducting oxypnictide (La$_{0.8}$Th$_{0.2}$OFeAs) was obtained by substituting the trivalent La ion with the tetravalent Th ions in LaOFeAs. This shows the highest transition temperature of 30.3 K in the lanthanum based oxypnictides synthesized at ambient pressure by electron doping. The temperature dependence of the penetration depth shows exponential behavior and the best fit is obtained for a s-wave pairing mechanism with a gap value of 1.4 meV. Thermoelectric power measurement indicates that the dominant carriers are like electrons. Both thermopower and resistivity studies indicate presence of strong electron –electron correlation as compared to the F-doped or O-deficient LaOFeAs. From magnetoresistance studies we found $H_{c2}(0)$ values of over 47 T that corresponds to a coherence length of 26 Å for this new superconductor.

## Acknowledgement:

AKG and SP thank DST, Govt. of India for financial support. JP and SJS thank CSIR and UGC, Govt. of India, respectively for fellowships. We thank AIF, JNU for the EDAX measurements.

## References:

1. Kamihara Y, Watanabe T, Hirano M and Hosono H 2008 J. *Am. Chem. Soc.* **130** 3296-97
2. Ma F and Lu Z-Y 2008 *Phys. Rev. B* **78** 033111
3. Dong J, Zhang H J, Xu G, Li Z, Li G, Hu W Z, Wu D, Chen G F, Dai X, Luo J L, Fang Z and Wang N L 2008 *Europhys. Lett*. **83** 27006.
4. Cruz C, Huang Q, Lynn J W, Li J, Ratcliff W, Zarestky J L, Mook H A, Chen G F, Luo J L, Wang N L and Dai P 2008 *Nature (London)* **453** 899-902
5. McGuire A M, Christianson A D, Sefat A S, Jin R, Payzant E A, Sales B C, Lumsden M D and Mandrus D, *arXiv: 0804.0796* (unpublished)
6. Chen G F, Li Z, Wu D, Li G, Hu W Z, Dong J, Zheng P, Luo J L and Wang N L 2008 *Phys. Rev. Lett.* **100** 247002 ; Prakash J, Singh S J, Patnaik S and Ganguli A K 2008 *Physica* C (In Print)
7. Chen X H, Wu T, Wu G, Liu R H, Chen H and Fang D F 2008 *Nature (London)* **453** 761-62
8. Ren Z -A, Yang J, Lu W, Yi W, Shen X-L, Li Z-C, Che G-C, Dong X-L, Sun L-L, Zhou F and Zhao Z-X 2008 *Europhys. Lett.* **82** 57002
9. Ren Z-A, Yang J, Lu W, Yi W, Che G-C, Dong X-L, Sun L-L and Zhao Z-X 2008 *Mater. Res. Innovations* **12** 105-06
10. Yang J, Li Z-C, Lu W, Yi W, Shen X-L, Ren Z-A, Che G-C, Dong X-L, Sun L-L, Zhou F and Zhao Z-X 2008 *Supercond. Sci. Technol.* **21** 082001-03
11. Takahashi H, Igawa K, Arii K, Kamihara Y, Hirano M and Hosono H 2008, *Nature(London)* **453** 376-78
12. Prakash J, Singh S, Samal S, Patnaik S and Ganguli A K 2008 *Europhys. Lett.* **84** 57003
13. Prakash J, Singh S J, Patnaik S and Ganguli A K 2008 *Solid State Commun.* **149** 181-83

14. Wang C, Li L, Chi S, Zhu Z, Ren Z, Li Y, Wang Y, Lin X, Luo Y, Jiang S, Xu X, Cao G and Xu Z 2008 *Europhys. Lett*. **83** 67006

15. Li L, Li Y, Ren Z, Luo Y, Lin X, He M, Tao Q, Zhu Z, Cao G and Xu Z 2008 *Phys. Rev. B* **78** 132506

16. Xu M, Chen F, He C, Ou H-W, Zhao J F and Feng D L 2008 *Chem. Mater.* **20** 7201-04

17. Sefat A S, Mcquire M A, Sales B C, Jin R, Howe J Y and Mandrus D 2008 *Phys. Rev. B* **77** 174503

18. Patnaik S, Singh K J and Budhani R C, *Rev. Sci. Instrum*. **70** 1494

19. Quebe P, Terbuchte L J and Jeitschko W 2000 *J. Alloys Compd.* **302** 70-74

20. Ren Z A, Che G-C, Dong X-L, Yang J, Lu W, Yi W, Shen X-L, Li Z-C, Sun L-L, Zhou F and Zhao Z-X 2008 *Europhys. Lett.* **83** 17002

21. Shan L, Wang Y, Zhu X, Mu G, Fang L and Wen H-H *arXiv: 0803.2405v2*(unpublished)

22. Carrington A and Manzano F 2003 *Physica C* **385** 205-14

23. Hunte F, Jaroszynski J, Gurevich A, Larbalestier D C, Jin R, Sefat A S, MacGuire M A, Sales B C, Christen D K and Mandrus D 2008 *Nature(London)* **453** 903-05

24. Gang M, Yu Z X, Lei F, Lei S, Cong R and Hu W H 2008 *Chin. Phys. Lett*. **25** 2221-24

25. Manzano F, Carrington A, Hussey N E, Lee S, Yamamoto A and Tajima S 2002 *Phys. Rev. Lett*. **88** 047002

26. Kawasaki S, Shimada K, Chen G F, Luo J L, Wang N L and Zheng G, *arXiv: 0810.1818* (unpublished)

27. Palstra T T M, Batlogg B, Schneemeyer L F and Waszczak J V 1988 *Phys. Rev. Lett.* **61** 1662-65

28. Patnaik S, Gurevich A, Kaushik S D, Bu S D, Choi J, Eom C B and Larbalestier D C 2004 *Phys. Rev. B* **70** 064503

29. Werthamer N R, Helfand E and Hohenberg P C 1966 *Phys. Rev.* **147** 295-302

30. Sasmal K, Lv B, Lorenz B, Guloy A M, Chen F, Xue Y Y and Chu C W 2008 *Phys. Rev. Lett*. **101** 107007

# Figure caption:

**Figure: 1. (i)** Powder X-ray diffraction patterns (XRD) of (a) $La_{0.9}Th_{0.1}OFeAs$ (b) $La_{0.8}Th_{0.2}OFeAs$ (c) $La_{0.9}OFeAs$ (d) $La_{0.8}OFeAs$ (e) $LaO_{0.9}FeAs$ and (f) $LaO_{0.8}FeAs$ sintered at 1180 C. The impurity phases are $ThO_2$ (*). **(ii)** EDX spectrum confirming the presence of La, Th, Fe, As and O.

**Figure: 2. (a)** The temperature dependence of resistivity ($\rho$) as a function of temperature for (a) $La_{0.8}OFeAs$ (b) $La_{0.9}OFeAs$ (c) $LaO_{0.8}FeAs$ (d) $LaO_{0.9}FeAs$ and (e) $La_{0.9}Th_{0.1}OFeAs$ **(b)** The variation of zero field resistivity ($\rho$) with respect to temperature in $La_{0.8}Th_{0.2}OFeAs$. Inset shows resistivity up to room temperature.

**Figure:3**. Low temperature (<9K) variation of penetration depth $\Delta\lambda$ (T) in polycrystalline $La_{0.8}Th_{0.2}OFeAs$. The red line and blue line show the exponential and power law ($T^2$) fitting respectively. Inset shows the temperature dependence of penetration depth upto 35 K attesting onset of bulk diamagnetism at $T_c$.

**Figure: 4.** Temperature dependence of the electrical resistivity of $La_{0.8}Th_{0.2}OFeAs$ superconductors under varying magnetic fields. Insets show temperature dependence of upper critical field (■) and irreversibility field (●) as a function of temperature.

**Figure: 5.** Temperature dependent thermoelectric power for the sample $La_{0.8}Th_{0.2}OFeAs$. Inset shows the in plane electrical resistivity of $La_{0.8}Th_{0.2}OFeAs$ as a function of temperature under zero field and fit of data for 31 K $\leq$ T $\leq$ 150 K to the form $\rho = \rho_0 + BT^2$.

**Figure 1**

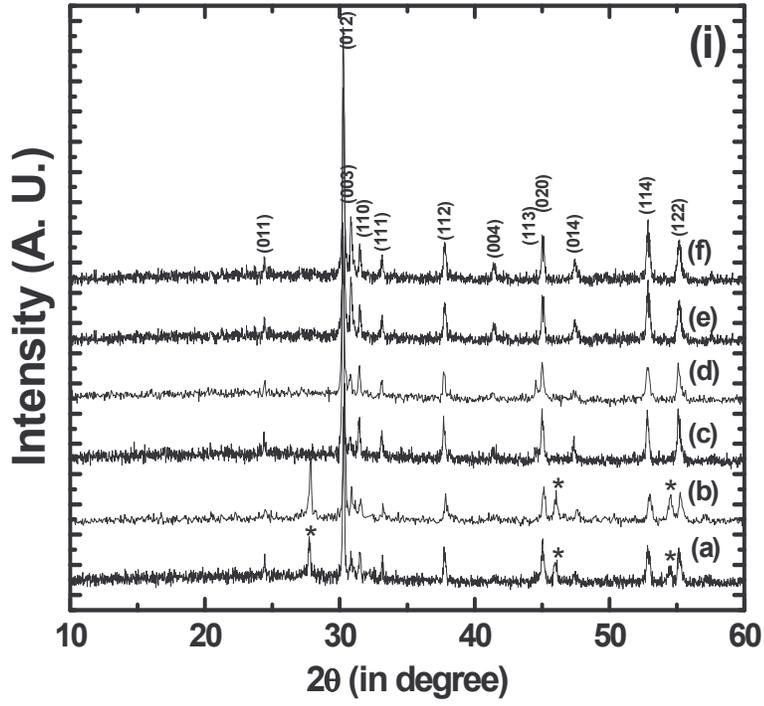

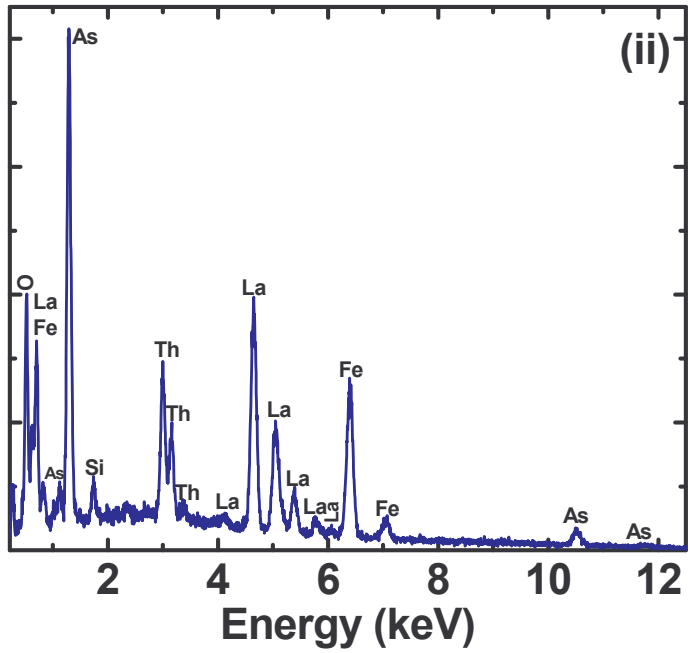

Figure 2

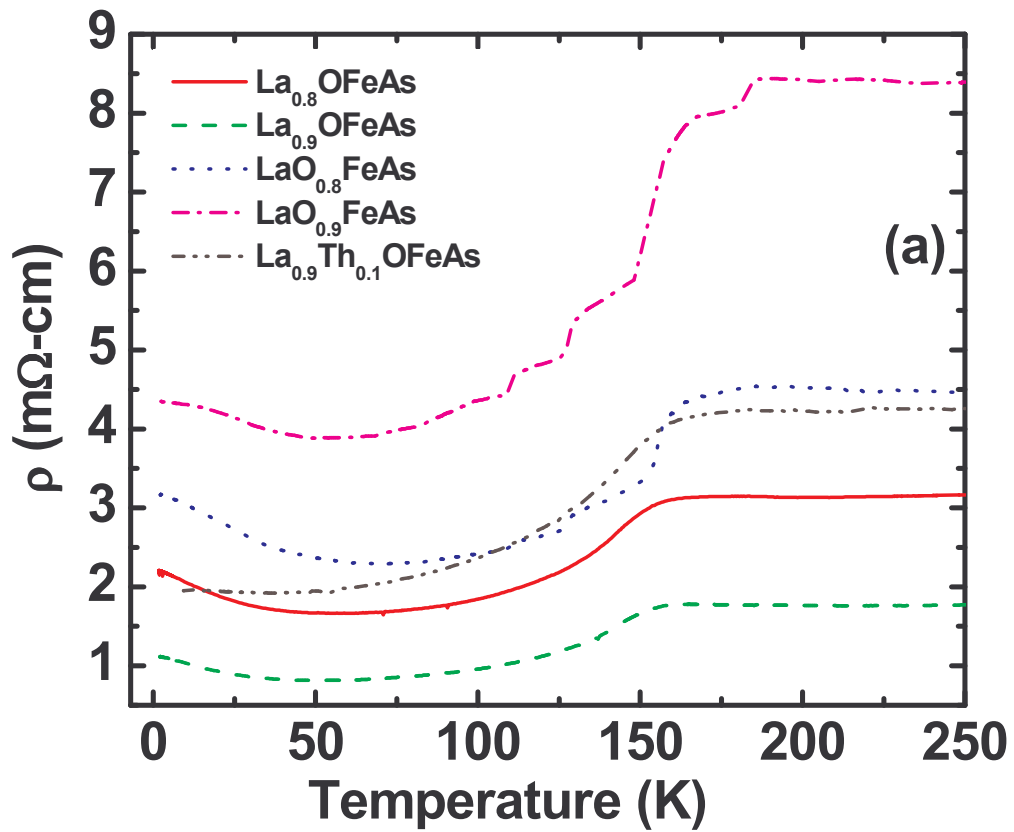

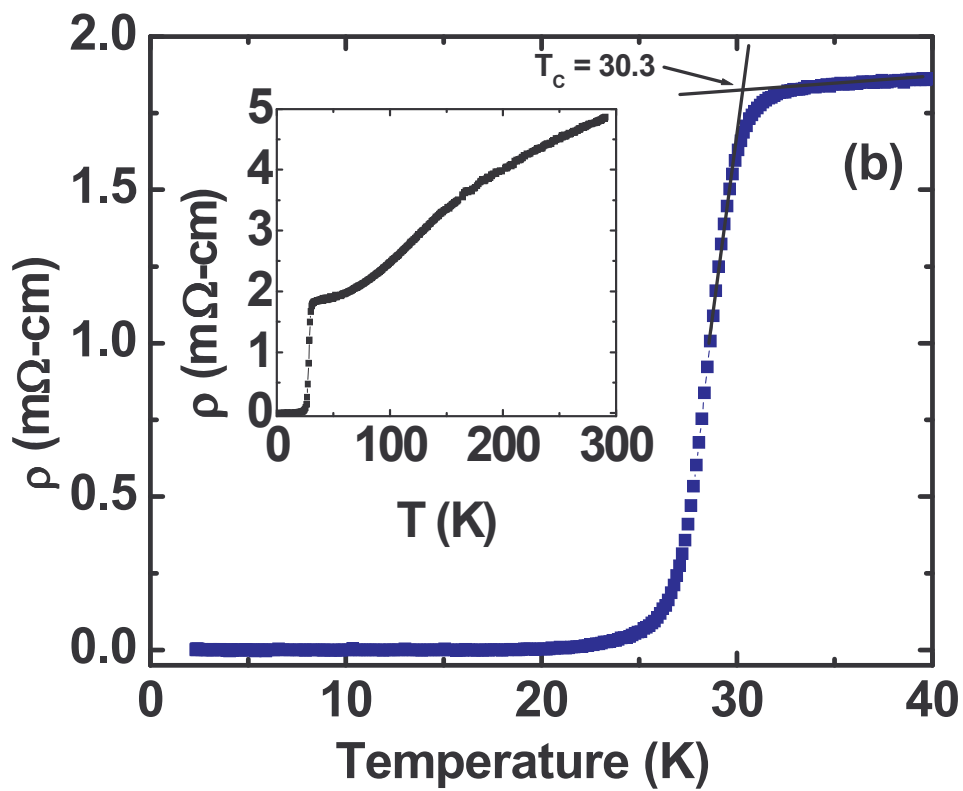

**Figure 3**

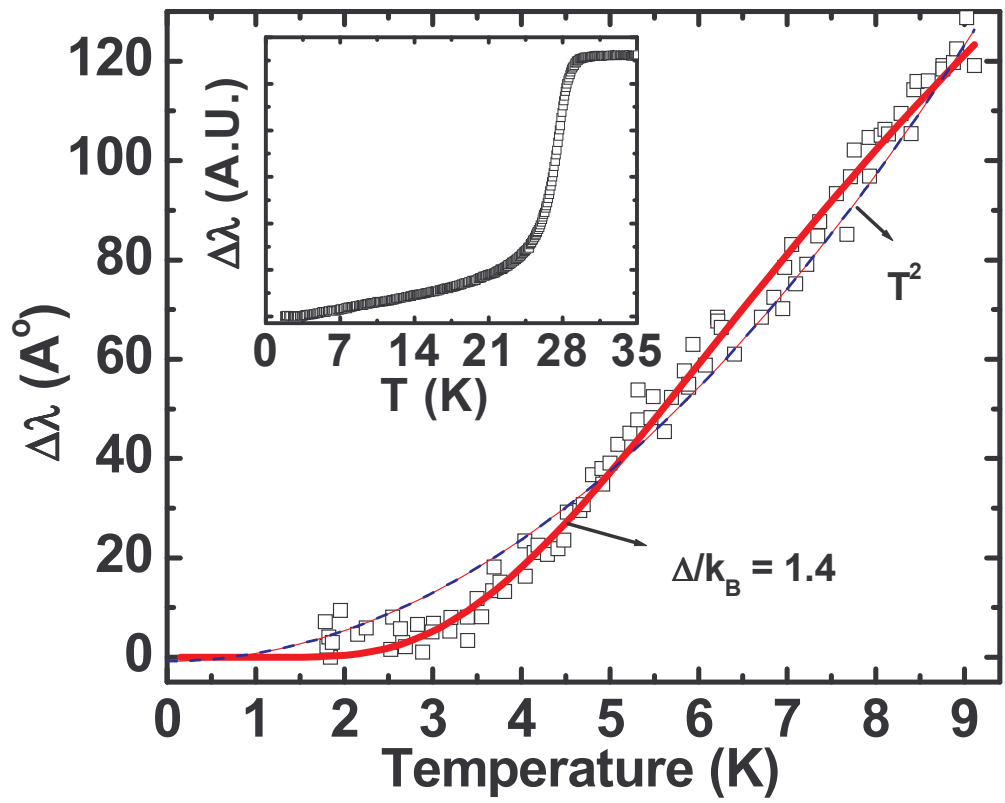

Figure 4

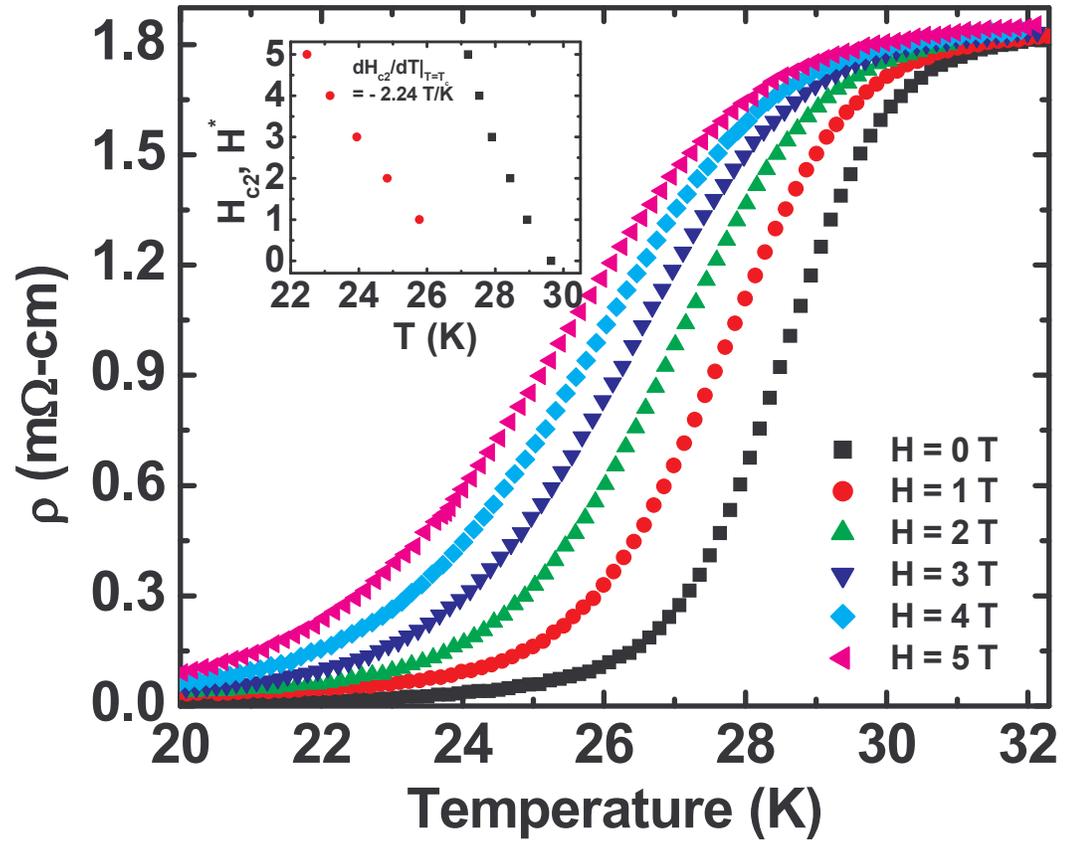

Figure 5

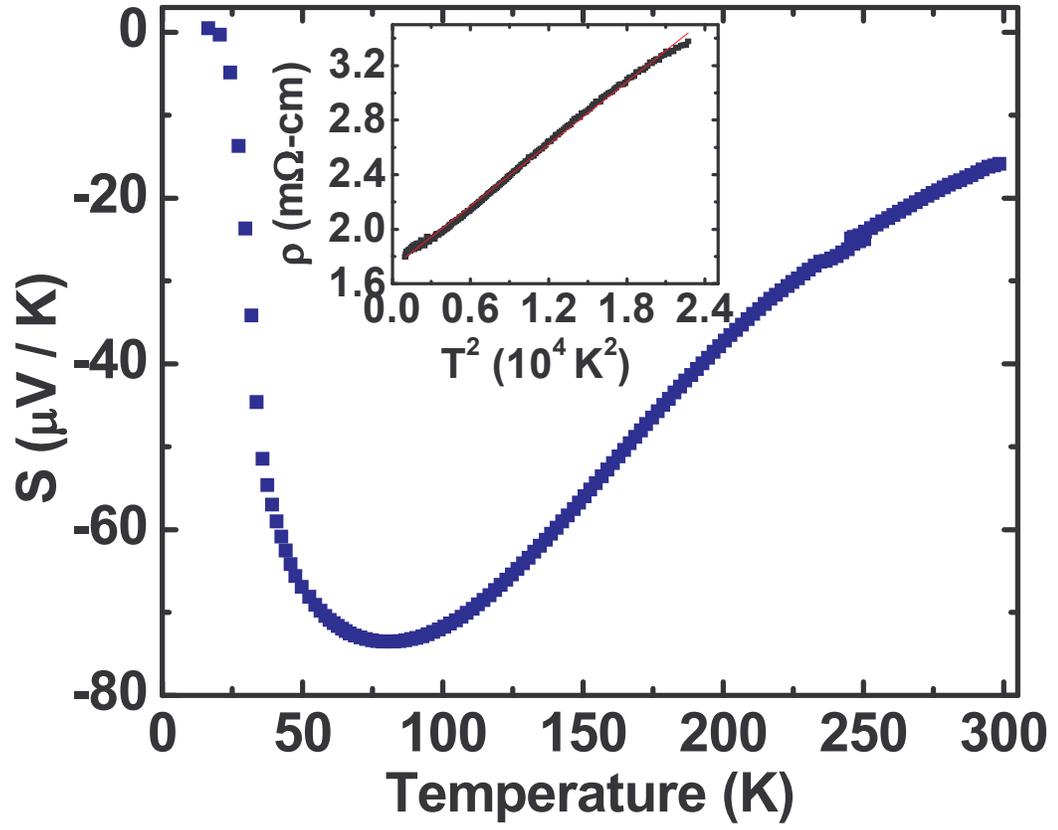